\documentclass[%
reprint,
showpacs,preprintnumbers,
 amsmath,amssymb,
 aps,
]{revtex4-1}

\usepackage{graphicx}
\usepackage{dcolumn}
\usepackage{bm}
\usepackage{subfigure} 
\usepackage{color,soul}
\usepackage{ulem}

\begin{document}


\title{Spreading law on a completely wettable spherical substrate: The energy balance approach
}

\author{Masao Iwamatsu}
\email{iwamatsu@ph.ns.tcu.ac.jp}
\affiliation{%
Department of Physics, Faculty of Liberal Arts and Sciences, Tokyo City University, Setagaya-ku, Tokyo 158-8557, Japan
}%



\date{\today}

\begin{abstract}
The spreading of a cap-shaped spherical droplet on a completely wettable spherical substrate is studied.  The non-equilibrium thermodynamic formulation is used to derive the thermodynamic driving force of spreading including the line-tension effect.  Then the energy balance approach is adopted to derive the evolution equation of the spreading droplet.  The time evolution of the contact angle $\theta$ of a droplet obeys a power law $\theta \sim t^{-\alpha}$ with the exponent $\alpha$, which is different from that derived from Tanner's law on a flat substrate.  Furthermore, the line tension must be positive to promote complete wetting on a spherical substrate, while it must be negative on a flat substrate.
\end{abstract}

\maketitle

\section{Introduction}
The spreading of a liquid droplet on a solid substrate is a complicated phenomena where many factors come into play.   However, the time evolution of the spreading of a liquid droplet on a flat solid surface can be usually described by simple universal power laws~\cite{Voinov1976,Tanner1979,Hervet1984,deGennes1985,Seaver1994,deRuijter2000,Daniel2006}.  The most famous law called Tanner's law describes the spreading of small non-volatile droplet on a completely wettable substrate.  This law has been derived theoretically using several different approaches~\cite{Voinov1976,Tanner1979,Hervet1984} and confirmed experimentally~\cite{Tanner1979,deRuijter1999,Rafai2004}. So far, however, most of theoretical as well as experimental work is confined to a droplet on a flat substrate.  Furthermore, the line-tension effect, which acts at the three-phase contact line and must play important role, has not been considered except for few theoretical works~\cite{Fan2006,Mechkov2009}.

In the present study, we will consider the problem of spreading of a cap-shaped spherical droplet of non-volatile Newtonian fluids~\cite{deGennes1985} gently placed on a spherical substrate, though our results might be applicable to non-Newtonian fluids as well~\cite{Rafai2004}.  The spreading on a spherical substrate is interesting because not only several experimental works has started to appear~\cite{Tao2011,Eral2011,Extrand2012} but also it has recently been revealed that the wetting behavior on a spherical substrate is totally different from that on a flat substrate, in particular, when the line tension is important~\cite{Iwamatsu2015,Iwamatsu2016a,Iwamatsu2016b}.  In fact, the effect of line tension on a spherical substrate is different from that on a flat substrate.  For example, the complete wetting state can be realized by {\it positive} line tension on a spherical substrate~\cite{Iwamatsu2016a}, while it can be realized by {\it negative} line tension on a flat substrate~\cite{Widom1995}. 

Although the magnitude of the line tension is believed to be small~\cite{Pompe2000,Wang2001,Checco2003,Schimmele2007,Berg2010} so that the size of the droplet must be nano-scale, there is some argument that the line tension could be a few order of magnitude larger~\cite{Herminghaus2006, Law2017} than it has bee predicted so far when the gravitation can be important.  If that is true, it could be possible to observe the line-tension effect on a micro- to millimeter scale droplets.  Note, however, that the experimental determination of line tension from the contact angle measurements is problematical because various effects such as the adsorption at the three-phase contact line~\cite{Ward2008}, the effect of substrate on the surface tension through the disjoining pressure~\cite{Napari2003,MacDowell2013}, and curvature-dependent surface tension~\cite{Tolman1949,Liu2013} will hinder the unique interpretation of experimental results to extract intrinsic line tension.  In the following, we will consider the spreading of a cap-shaped spherical droplet on a spherical substrate using the energy-balance approach~\cite{Hervet1984,deGennes1985,Daniel2006} which can easily include the line-tension effect~\cite{Mechkov2009}.

\section{\label{sec:sec2}Effect of line tension on thermodynamic driving force}

\begin{figure}[htbp]
\begin{center}
\includegraphics[width=0.80\linewidth]{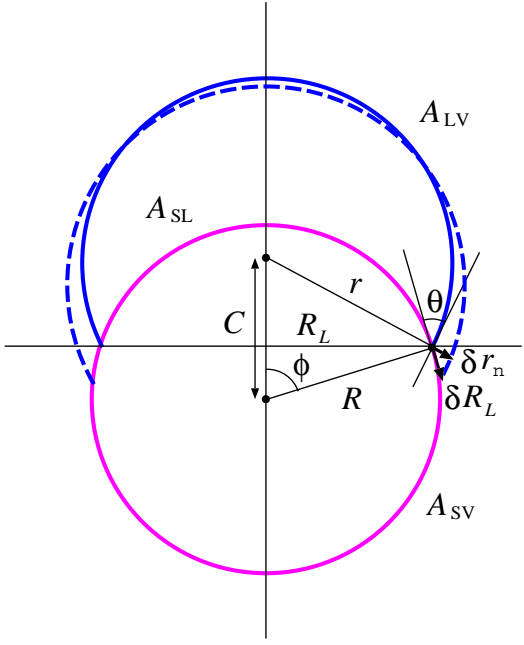}
\caption{
A spherical liquid droplet on a convex spherical substrate.  The centers of the droplet with radius $r$ and that of the spherical substrate with radius $R$ are separated by a distance $C$.  The radius of the contact line is denoted by $R_{\rm L}$ and the dynamic contact angle is denoted by $\theta$. Note that the three-phase contact line passes through the equator when the half of the central angle becomes $\phi=90^{\circ}$  and the contact line moves from the upper hemisphere to lower hemisphere on the sphere.  A small distortion of a spherical cap induces the displacement of radius $\delta r_{\rm n}$ and that of contact line $\delta R_{\rm L}$. }
\label{fig:1}
\end{center}
\end{figure}

The non-equilibrium thermodynamics~\cite{Fan2006} will be used to derive the thermodynamic driving force for spreading.  Suppose, the droplet is a spherical cap of radius $r$ with the non-equilibrium dynamic contact angle $\theta$ on a spherical substrate of radius $R$ as shown in Fig.~\ref{fig:1}.  The Gibbs free energy $G$ consists of the interface free energy $F$ and the work supplied by the external action $w$:
\begin{equation}
G=F-w
\label{eq:SP1}
\end{equation}
with
\begin{equation}
F=\sigma_{\rm LV}A_{\rm LV}+\sigma_{\rm SV}A_{\rm SV}+\sigma_{\rm SL}A_{\rm SL}+\tau L
\label{eq:SP2}
\end{equation}
where $\gamma_{ij}$ and $A_{ij}$ are the surface tension and surface area of liquid-vapor (LV), solid-vapor (SV) and solid-liquid (SL) interfaces.  The last term is the contribution form the line tension $\tau$ with the three-phase contact line length $L$.  

From Eq.~(\ref{eq:SP2}), the free energy change by a small distortion of a spherical cap (Fig.~\ref{fig:1}) can be expressed by
\begin{equation}
\delta G=\sigma_{\rm LV}\delta A_{\rm LV}+\sigma_{\rm SV}\delta A_{\rm SV}+\sigma_{\rm SL}\delta A_{\rm SL}+\tau\delta L-\delta w.
\label{eq:SP3}
\end{equation}
A small distortion induces the displacement of radius $\delta r_{\rm n}$ and that of contact line $\delta R_{\rm L}$ of the droplet.  Therefore, there are two contributions to the increase of liquid-vapor surface area $\delta A_{\rm LV}=\delta A_{\rm LV1}+\delta A_{\rm LV2}$.  The first contribution is given by (see Fig.~\ref{fig:1}) 
\begin{equation}
\delta A_{\rm LV1}=\int\int_{\rm S_{\rm LV}} \kappa \delta r_{\rm n} dA,
\label{eq:SP4}
\end{equation}
where the integration is over the liquid-vapor surface area $\rm S_{\rm LV}$, and
\begin{equation}
\kappa=\frac{2}{r}
\label{eq:SP5}
\end{equation}
is the curvature of the spherical liquid-vapor interface with radius $r$.  The displacement $\delta R_{\rm L}$ of the triple line in Fig.~\ref{fig:1} by spreading will also increase the liquid-vapor surface area, which is the second contribution given by
\begin{equation}
\delta A_{\rm LV2}=\oint_{\rm L} \delta R_{\rm L}\cos\theta dl,
\label{eq:SP6}
\end{equation}
where the integration is along the three-phase contact line L.  The increase of the solid-liquid surface area is the decrease of the solid-vapor surface area, and is given by
\begin{equation}
\delta A_{\rm SL}=\oint_{\rm L} \delta R_{\rm L}dl=-\delta A_{\rm SV}.
\label{eq:SP7}
\end{equation}
The radius of the triple line increases by the amount $\delta R_{\rm L}\cos\phi$, therefore
\begin{equation}
\delta L=\oint_{\rm L}\frac{\delta R_{\rm L}\cos\phi}{R_{\rm L}}dl,
\label{eq:SP8}
\end{equation}
where $R_{\rm L}=R\sin\phi$ is the radius of the triple line and $\phi$ is half of the central angle corresponding to the contact arc of the solid surface (Fig.~\ref{fig:1}). Using the work done by the Laplace pressure $\Delta p$, the increase of the work $\delta w$ supplied by external action is written as
\begin{equation}
\delta w=\int\int_{S_{\rm LV}} \Delta p \delta r_{\rm n} dA.
\label{eq:SP9}
\end{equation}
Therefore, the free energy change $\delta G$ in Eq.~(\ref{eq:SP3}) is givey by
\begin{eqnarray}
\delta G&=&\sigma_{\rm LV}\int\int_{S_{\rm LV}} \kappa \delta r_{\rm n} dA+\sigma_{\rm LV}\oint_{L} \delta R_{\rm L}\cos\theta dl
\nonumber \\
&&-\sigma_{\rm SV}\oint_{L} \delta R_{\rm L}dl+\sigma_{\rm SL}\oint_{L} \delta R_{\rm L}dl
\nonumber \\
&&+\tau\oint_{L}\frac{\delta R_{\rm L}\cos\phi}{R_{\rm L}}dl-\int\int_{S_{\rm LV}} \Delta p \delta r_{\rm n} dA.
\label{eq:SP10}
\end{eqnarray}

Defining the thermodynamic driving forces $f_{\rm S}$ and $f_{\rm L}$ through
\begin{equation}
\oint_{L}f_{\rm L}\delta R_{\rm L}dl+\int\int_{S_{\rm LV}} f_{\rm S} \delta r_{\rm n} dA=-\delta G,
\label{eq:SP11}
\end{equation}
we obtain
\begin{eqnarray}
f_{\rm L}&=&\sigma_{\rm SV}-\sigma_{\rm SL}-\sigma_{\rm LV}\cos\theta-\frac{\tau\cos\phi}{R_{\rm L}},
\label{eq:SP12} \\
f_{\rm S}&=&\Delta p-\sigma_{\rm LV}\kappa.
\label{eq:SP13}
\end{eqnarray}
Using the spreading parameter $S$ defined by
\begin{equation}
S=\sigma_{\rm SV}-\sigma_{\rm SL}-\sigma_{\rm LV},
\label{eq:SP14}
\end{equation}
Eq.~(\ref{eq:SP12}) can be written as
\begin{equation}
f_{\rm L}=
S+\sigma_{\rm LV}\left(1-\cos\theta\right)-\frac{\tau}{R\tan\phi}.
\label{eq:SP15}
\end{equation}
At the thermodynamic equilibrium $f_{\rm L}=f_{\rm S}=0$, Eqs.~(\ref{eq:SP12}) and (\ref{eq:SP13}) become
\begin{eqnarray}
0&=&S+\sigma_{\rm LV}\left(1-\cos\theta\right)-\frac{\tau}{R\tan\phi},
\label{eq:SP16} \\
0&=&\Delta p-\sigma_{\rm LV}\kappa,
\label{eq:SP17}
\end{eqnarray}
which are the generalized Young's formula and Young-Laplace formula, respectively. When the substrate is incompletely-wettable ($S<0$) and is characterized by the Young's contact angle $\theta_{\rm Y}$ defined through 
\begin{equation}
\sigma_{\rm SV}-\sigma_{\rm SL}=\sigma_{\rm LV}\cos\theta_{\rm Y},
\label{eq:SP18}
\end{equation}
Eq.~(\ref{eq:SP16}) becomes
\begin{equation}
\sigma_{\rm LV}\left(\cos\theta_{\rm Y}-\cos\theta\right)-\frac{\tau}{R\tan\phi}=0,
\label{eq:SP19}
\end{equation}
because $S=\sigma_{\rm LV}\left(\cos\theta_{\rm Y}-1\right)<0$. Equation (\ref{eq:SP19}) will determine the equilibrium contact angle $\theta_{\rm e}$ which includes the line-tension effect when the substrate is characterized by the Young's contact angle $\theta_{\rm Y}$.  However, because we pay most attention to the completely-wettable substrate with $S\ge 0$, we will use Eqs.~(\ref{eq:SP15}) and (\ref{eq:SP16}) instead of Eq.~(\ref{eq:SP19}).

Using the geometric relations (Fig.~\ref{fig:1})
\begin{eqnarray}
C\cos\phi &=& R-r\cos\theta,
\nonumber \\
C\sin\phi &=& r\sin\theta,
\label{eq:SP20}
\end{eqnarray}
and the size parameter of the droplet
\begin{equation}
\rho=\frac{r}{R},
\label{eq:SP21}
\end{equation}
relative to the size of the spherical substrate, we obtain
\begin{equation}
\tan\phi=\frac{\rho\sin\theta}{1-\rho\cos\theta}.
\label{eq:SP22}
\end{equation}
Then, the thermodynamic force $f_{\rm L}$ in Eq.~(\ref{eq:SP15}) is written as
\begin{equation}
f_{\rm L}=S+\sigma_{\rm LV}\left(1-\cos\theta\right)-\frac{1-\rho\cos\theta}{r\sin\theta}\tau
\label{eq:SP23}
\end{equation}
using the dynamic contact angle $\theta$. The line-tension contribution vanishes and changes its sign at $\phi=\phi_{\rm c}=\pi/2$ from Eq.~(\ref{eq:SP15}) or at $\theta=\theta_{\rm c}$ defined by $R-r\cos\theta_{\rm c}=0$ from Eq.~(\ref{eq:SP23}), which gives the critical contact angle $\theta_{\rm c}$ given by
\begin{equation}
\cos\theta_{\rm c} = \frac{R}{r}=\frac{1}{\rho},
\label{eq:SP24}
\end{equation}
where the three-phase contact line passes through the equator of the substrate.

\begin{figure}[htbp]
\begin{center}
\includegraphics[width=1.0\linewidth]{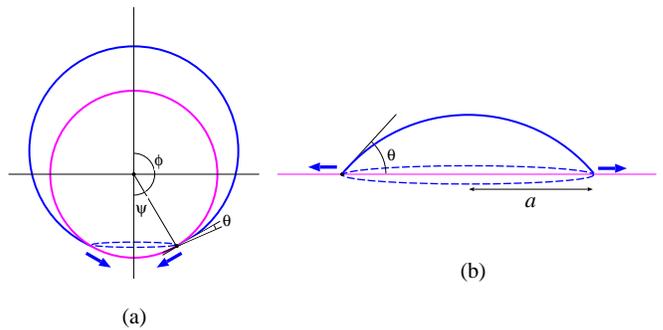}
\caption{
(a) A spreading droplet on a spherical substrate.   The three-phase contact line shrinks towards the south pole of the substrate to realize complete wetting state.  (b) A spherical droplet on a flat substrate.  In this case, the three-phase contact line expands toward infinity to realize the complete wetting state.  Therefore, the line tension on a spherical substrate plays opposite role to that on a flat substrate.
 }
\label{fig:2}
\end{center}
\end{figure}

During the spreading with dynamic contact angle $\theta$, a positive line tension ($\tau>0$) will decelerate the spreading of the three-phase contact when the three phase contact line is on the upper hemisphere ($\phi<\pi/2$) from Eq.~(\ref{eq:SP15}).  However, once the three-phase contact line crosses the equator and moves into the lower hemisphere ($\phi>\pi/2$), the positive line tension will accelerate the spreading.  This fact can be easily understood as the spreading on the upper hemisphere accompanies the expansion of the contact line (circle), while the complete wetting on the lower hemisphere is accomplished by the shrinkage of the contact circle (Fig.~\ref{fig:2}(a)).  In contrast, the positive line tension ($\tau>0$) will always decelerate the spreading on a flat substrate~\cite{Mechkov2009} as the contact line will always expand towards the complete wetting state (Fig.~\ref{fig:2}(b)).

The size parameter $\rho$ or the droplet radius $r$ is not a constant but is a function of the dynamic contact angle $\theta$, since the droplet volume $V$ is fixed for a non-volatile liquid.  The droplet volume $V$ is given by~\cite{Iwamatsu2015}
\begin{equation}
V
=\left(\frac{4\pi}{3}R^{3}\right)\frac{(\zeta-1+\rho)^{2}\left[3\left(1+\rho\right)^{2}-2\zeta\left(1-\rho\right)-\zeta^{2}\right]}{16\zeta},
\label{eq:SP25}
\end{equation}
with
\begin{equation}
\zeta=\sqrt{1+\rho^{2}-2\rho\cos\theta}.
\label{eq:SP26}
\end{equation}
Since the droplet volume is fixed at
\begin{equation}
V_{0}=\frac{4\pi}{3}R^{3}\left(\rho_{0}^{3}-1\right),
\label{eq:SP27}
\end{equation}
where $\rho_{0}=r_{0}/R>1$ is the size parameter when the droplet completely spread over the spherical substrate ($\theta=0$), we obtain
\begin{equation}
\rho\simeq \rho_{0} -\frac{\rho_{0}}{16\left(\rho_{0}-1\right)^{3}}\theta^{4}
\label{eq:SP28}
\end{equation}
by expanding Eq.~(\ref{eq:SP25}) and $\rho$ by $\theta$.  Then, Eq.~(\ref{eq:SP22}) is approximated by
\begin{equation}
\tan\phi\rightarrow \frac{\rho_{0}}{1-\rho_{0}}\theta.
\label{eq:SP29}
\end{equation}
when $\theta\rightarrow 0$ and $\phi\rightarrow \pi$.  Note that we cannot extrapolate our results of the spherical substrate to a flat substrate by taking the limit $R\rightarrow \infty$ or $\rho_{0}\rightarrow 0$ in Eqs. (27) to (29) because we alwasy have $\rho_{0}>1$ as the droplet volume is fixed.  In order to wet a spherical substrate completely, the surface area of the spreading droplet must always be larger than that of the substrate.  In contrast, the surface area of the droplet must always be smaller that that of an infinite flat substrate.  This topological difference already implies that the spreading law on a spherical substrate is different from that on a flat substrate.

\section{\label{sec:sec2}Spreading using the energy balance approach}

Once we found the thermodynamic driving force $f_{\rm L}$ in Eqs.~(\ref{eq:SP15}) and (\ref{eq:SP23}) for spreading, we can formulate the spreading problem of a droplet on a completely wettable ($S>0$) spherical substrate.  We will follow closely the energy balance approach originally formulated by de Genne~\cite{deGennes1985} and extended to include the line tension on a flat substrate by Mechkov et al.~\cite{Mechkov2009}.

Suppose the free energy of the contact line per unit length is given by $g_{\rm L}$ and the free energy dissipation at the contact line per unit length per unit time is $W$, then the free energy of the contact line is given by $G_{\rm L}=2\pi R_{L}g_{\rm L}$.  Since the contact line length is $R_{\rm L}=2\pi R\sin\phi$, the free energy balance is expressed as
\begin{equation}
-\frac{d}{dt}\left(2\pi R\sin\phi g_{\rm L}\right)=\left(2\pi R\sin\phi\right) W.
\label{eq:SP30}
\end{equation}
On the other hand, the free energy dissipation due to the action of the thermodynamic force $f_{\rm L}$ can be written as
\begin{equation}
-\frac{d}{dt}\left(2\pi R\sin\phi g_{\rm L}\right)=\left(2\pi R\sin\phi\right) vf_{\rm L}
\label{eq:SP31}
\end{equation}
using the velocity $v$ of the contact line.  Since the right-hand side of Eqs.~(\ref{eq:SP30}) and (\ref{eq:SP31}) must be equal, we have
\begin{equation}
W=v f_{\rm L}=v\left(S+\sigma_{\rm LV}\left(1-\cos\theta\right)-\frac{\tau}{R\tan\phi}\right),
\label{eq:SP32}
\end{equation}
which is the starting point of the energy balance approach.

According to de Gennes~\cite{deGennes1985}, the dissipation $W$ can be divided into three contributions
\begin{equation}
W=W_{\rm drop} + W_{\rm film} + W_{\rm precursor},
\label{eq:SP33}
\end{equation}
where $W_{\rm drop}$ is the dissipation in the wedge of the drop, $W_{\rm film}$ is the dissipation in the wetting film, and $W_{\rm precursor}$ is the dissipation in the precursor film.  Hervet and de Gennes~\cite{Hervet1984,deGennes1985} proved that
\begin{equation}
W_{\rm film}=vS,
\label{eq:SP34}
\end{equation}
which might be questionable on a spherical substrate as the film cannot exist when the droplet completely wet the substrate (Fig.~\ref{fig:2}(a)), while it can exist on an infinite flat substrate (Fig.~\ref{fig:2}(b)).  Since we are interested in the spreading process before complete wetting, we will continue to adopt Eq.~(\ref{eq:SP34}) on a spherical substrate.

Furthermore, the contribution of the precursor film $W_{\rm precursor}$ is negligible~\cite{deGennes1985,Mechkov2009}.  Then, Eq.~(\ref{eq:SP32}) becomes
\begin{equation}
W_{\rm drop} = v\left(\sigma_{\rm LV}\left(1-\cos\theta\right)-\frac{\tau}{R\tan\phi}\right),
\label{eq:SP35}
\end{equation}
where we consider only the hydrodynamic viscous dissipation $W_{\rm drop}$ inside the drop in Eq.~(\ref{eq:SP33}).  Using the wedge approximation, the dissipation in the bulk drop is given by~\cite{deGennes1985,Daniel2006,Mechkov2009}
\begin{equation}
W_{\rm drop}\simeq \frac{3\eta v^{2}}{\theta}\ln\left|\frac{x_{\rm max}}{x_{\rm min}}\right|=\frac{\kappa\eta v^{2}}{\theta}
\label{eq:SP36}
\end{equation}
where $\eta$ is the viscosity of the liquid, $x_{\rm max}$ and $x_{\rm min}$ are the cutoff length when the three-phase contact area is approximated by a wedge.  Here, we introduced constant $\kappa=3\ln\left|x_{\rm max}/x_{\rm min}\right|$. Therefore, Eq.~(\ref{eq:SP35}) is written as
\begin{equation}
\frac{\kappa\eta}{\theta}v^2= v\left(\sigma_{\rm LV}\left(1-\cos\theta\right)-\frac{\tau}{R\tan\phi}\right).
\label{eq:SP37}
\end{equation}
We now introduce the angle $\psi$ defined by (Fig.~\ref{fig:2}(a))
\begin{equation}
\psi=\pi-\phi,
\label{eq:SP38}
\end{equation}
and consider the spreading to a complete wetting state $\theta\rightarrow 0$ and $\psi\rightarrow 0$.  Then, Eq.~(\ref{eq:SP29}) gives
\begin{equation}
\tan\phi \simeq -\psi=-\frac{\rho_{0}}{\rho_{0}-1}\theta
\label{eq:SP39}
\end{equation}
and Eq.~(\ref{eq:SP37}) becomes
\begin{equation}
\kappa\frac{\eta v}{\sigma_{\rm LV}}=\frac{1}{2}\theta^{3}+\frac{\rho_{0}-1}{\rho_{0}}\tilde{\tau},
\label{eq:SP40}
\end{equation}
where
\begin{equation}
\tilde{\tau}=\frac{\tau}{\sigma_{\rm LV}R}
\label{eq:SP41}
\end{equation}
is the scaled line tension relative to the liquid-vapor surface tension $\sigma_{\rm LV}$.

When the line tension $\tau$ can be neglected ($\tau=0$), we recover the standard equation of spreading given by~\cite{Voinov1976,Tanner1979,Seaver1994}
\begin{equation}
\theta^{3}= 2\kappa {\rm Ca}
\label{eq:SP42}
\end{equation}
from Eq.~(\ref{eq:SP40}), where
\begin{equation}
{\rm Ca}=\frac{\eta v}{\sigma_{\rm LV}}
\label{eq:SP43}
\end{equation}
is known as the capillary number. In contrast to the spreading on a flat substrate~\cite{Mechkov2009}, a positive line tension $\tau>0$ always accelerate the spreading from Eq.~(\ref{eq:SP40}).

Because Eq.~(42) and, therefore, Eq.~(40) are derived from Eq.~(36) which is based on the wedge approximation~\cite{Hervet1984,deGennes1985}, their validity might be questionable when the dynamic contact angle $\theta$ becomes small and the droplet and the spherical substrate have similar curvatures. Then, the droplet near the contact line looks more like a thin-film or a slab rather than a wedge.  In fact, an equation similar to Eq.~(36) can be derived by simplifying a spherical droplet by a thin-film or a disk~\cite{Seaver1994,deRuijter1999,Daniel2006}. Therefore, we will continue to use Eq.~(40) and (42) to discuss the spreading on a spherical substrate.

Since the velocity of the contact line on a sphere is given by
\begin{equation}
v=\frac{d}{dt}R\phi=-R\dot{\psi}=-R\frac{\rho_{0}}{\rho_{0}-1}\dot{\theta},
\label{eq:SP44}
\end{equation}
and Eq.~(\ref{eq:SP40}) will be transformed into the evolution equation for the dynamic contact angle $\theta$:
\begin{equation}
\dot{\theta}=-\frac{\sigma_{\rm LV}}{\kappa\eta R}\left(\frac{\rho_{0}-1}{\rho_{0}}\right)
\left(\frac{1}{2}\theta^{3}+\frac{\rho_{0}-1}{\rho_{0}}\tilde{\tau}\right).
\label{eq:SP45}
\end{equation}

When the line tension can be neglected ($\tau=0$), the evolution of the dynamic contact angle follow the evolution law
\begin{equation}
\theta \propto \psi \propto t^{-1/2},
\label{eq:SP46}
\end{equation}
and 
\begin{equation}
v  \propto t^{-3/2},
\label{eq:SP47}
\end{equation}
from Eqs.~(\ref{eq:SP44}) and (\ref{eq:SP45}).

On the other hand, when the line tension is {\it positive} ($\tau>0$) and dominant in Eq.~(\ref{eq:SP45}), we have
\begin{equation}
\theta \propto \psi \propto t_{0}-t
\label{eq:SP48}
\end{equation}
and
\begin{equation}
v \propto {\rm constant},
\label{eq:SP49}
\end{equation}
where $t_{0}$ is the time when the spreading will be completed and the droplet will enclose the spherical substrate, which is determined by the initial position (initial contact angle $\theta$) of the spreading front. This is the only scenario that can be deduced mathematically from our macroscopic model since the half of the central angle $\phi$ is always related to the dynamic contact angle $\theta$ through Eq.~(22). Then the dynamic contact angle $\theta$ approaches 0$^{\circ}$ ($\theta\rightarrow 0^{\circ}$) as the contact circle shrinks and disappears ($\phi\rightarrow \pi$). Microscopically, however, one may imagine that the advancement of spreading front is sufficiently fast that the contact circle is collapsing at a finite contact angle.  Then, the thickness of the wetting film at the south pole ($\phi=\pi$) of the spherical substrate substrate (Fig.~2(a)) changes discontinuously. In order to discuss such a microscopic process, we have to consider the microscopic model which includes the disjoining pressure and precursor film~\cite{deGennes1985,Mechkov2009}. When the line tension is negative ($\tau<0$), the droplet cannot spread over whole substrate since it is energetically unfavorable to eliminate the three-phase contact line.  In order to achieve the complete wetting, a {\it positive} line tension is necessary on a spherical substrate.

So far, we have considered a completely-wettable spherical substrate with $S>0$ or  $\theta_{\rm Y}=0^{\circ}$. When the line tension $\tau$ is positive and is sufficiently large on a partially-wettable substrate with $S<0$ or $\theta_{\rm Y}>0$,  the equilibrium morphology with the lowest free energy can be a complete-wetting state with the contact angle $\theta_{\rm e}=0^{\circ}$~\cite{Iwamatsu2016a}.  Then, the partially-wettable substrate turn into a completely wettable substrate by the action of line tension, and Eqs.~(48) and (49) become applicable.  In such a case, however, the initial position (initial contact angle $\theta$) of spreading front must start from the position where the free-energy barrier from a metastable state is already crossed~\cite{Iwamatsu2016a}. 

Finally, we recapture the result for a flat substrate~\cite{Voinov1976,Tanner1979,deGennes1985,Daniel2006} for the sake of comparison.  On a flat substrate, Eq.~(\ref{eq:SP40}) is written as~\cite{Mechkov2009}
\begin{equation}
\kappa\frac{\eta v}{\sigma_{\rm LV}}=\frac{1}{2}\theta^{3}-\frac{\tau}{\sigma_{\rm LV}a}\theta
\label{eq:SP50}
\end{equation}
where $a$ is the base radius of the droplet (Fig.~\ref{fig:2}(b)).  The droplet volume $V_{0}$ is proportional to the dynamic contact angle $\theta$ through
\begin{equation}
V_{0}\simeq \frac{\pi}{4}a^{3}\theta.
\label{eq:SP51}
\end{equation}
When the line tension can be neglected ($\tau=0$), Eq.~(\ref{eq:SP37}) together with $v=\dot{a}$ leads to the evolution equation of the base radius $a$, from which the well-know Tanner's law 
\begin{equation}
a \propto t^{1/10},
\label{eq:SP52}
\end{equation}
and
\begin{equation}
v \propto t^{-9/10}.
\label{eq:SP53}
\end{equation}
are derived.  The evolution law of the dynamic contact angle on a flat substrate is given by~\cite{deRuijter1999}
\begin{equation}
\theta \propto t^{-3/10},
\label{eq:SP54}
\end{equation}
which is different from Eq.~(\ref{eq:SP46}) on a spherical substrate.  

On the other hand, when the line tension is {\it negative} ($\tau<0$) rather than positive and dominant in Eq.~(\ref{eq:SP40}), we have~\cite{Mechkov2009}
\begin{eqnarray}
a &\propto& t^{1/5},
\label{eq:SP55a} \\
v &\propto& t^{-4/5},
\label{eq:SP55b}
\end{eqnarray}
and
\begin{equation}
\theta \propto t^{-3/5}.
\label{eq:SP56}
\end{equation}
which, again, is different from Eq.~(\ref{eq:SP48}) on a spherical substrate.  The scaling rule in Eq.~(\ref{eq:SP55a}) has been observed in the spreading of nematic crystals~\cite{Mechkov2009,Poulard2005,Poulard2006}.  In order to achieve the complete-wetting, a {\it negative} line tension is necessary on a flat substrate.

\section{\label{sec:sec5} Conclusion}
In the present study, we considered the problem of spreading of a cap-shaped spherical droplet on a spherical substrate using the energy balance approach. We found a scaling rules of the time evolution of the dynamic contact angle on a completely wettable spherical substrate. This new scaling rule is different from well-know Tanner's law~\cite{Tanner1979, deRuijter1999} on a flat substrate.  Experimental attempts to verify this scaling rule will be interesting.

The effect of line tension on the spreading on a spherical substrate is considered and the result is, again, different from that of a flat substrate.  Most notably, a {\it positive} line tension is necessary~\cite{Iwamatsu2016a} to realize complete wetting on a spherical substrate, while a {\it negative} line tension is necessary on a flat substrate.  The scaling rule of dynamics on a spherical substrate is also different from that on a flat substrate.  Even though the magnitude of the line tension has been believed to be small, a gravitation assisted enhancement of the line tension~\cite{Herminghaus2006,Law2017} would make it possible to observe line tension effect even in macroscopic droplets.

Although we consider the spreading of simple liquids on a hydrophilic substrate, the spreading of surfactant solutions on a hydrophilic flat substrate is known to show the non-isotropic fingering instabilities called Marangoni effect~\cite{Marmur1981,Hamaraoui2004,Craster2009} due to the surface tension gradient.  These instabilities are closely related to the general instabilities of the growing front due to diffusion~\cite{Mullins1963,Meakin1993}.  The fingering instabilities on a spherical surface will be an interesting subject to study.

Finally, we notice that the spreading on completely-wettable spherical substrate involves topological phase transition since the topology of the wetting film changes from a hollow to a spherical surface which enclose the spherical substrate.

\begin{acknowledgments}
This work was partially supported under a project for strategic advancement of research infrastructure for private universities, 2015-2020, operated by MEXT, Japan. The author is grateful to Professor Siegfried Dietrich (Max-Planck Institute for Intelligent Systems, Stuttgart) for sending him a useful material on line tension.
\end{acknowledgments}




\end{document}